\documentclass[aps,prl,twocolumn,showpacs]{revtex4}

\usepackage{amssymb}
\usepackage{graphicx}
\usepackage{epsfig}

\begin{document}

\date{\today}

\title{Instanton analysis of hysteresis in the 3D Random-Field Ising Model}
\author{Markus M\"uller and Alessandro Silva}
\affiliation{Department of Physics, Rutgers University,
Piscataway, New Jersey 08854} 

\begin{abstract}
We study the magnetic hysteresis in the random-field Ising model in
3D. We discuss the disorder dependence of the coercive field
$H_c$, and obtain an analytical description of the smooth part of
the hysteresis below and above $H_c$, by identifying the disorder
configurations (instantons) that are the most probable to trigger
local avalanches. We estimate the critical disorder strength at
which the hysteresis curve becomes continuous. From an
instanton analysis at zero field we obtain a description of local
two-level systems in the ferromagnetic phase.
\end{abstract}

\maketitle

 In disordered systems whose pure counterparts undergo a first
order phase transition as an external parameter (e.g., field or
pressure) is tuned, the presence of randomness leads to
interesting hysteretic phenomena~\cite{Sethnareview}. Those arise
from local modifications of the spinodal point due to disorder, as
well as from the presence of many metastable states. The simplest
system exhibiting such phenomena is the random field Ising model
(RFIM),
often studied as a representative model for hysteresis in the
presence of random fields
\cite{Berger00,Belanger02,Marcos03,Goh87,WongChan90,Tulimieri99}.

In the RFIM, the gradual transition between states of negative and
positive magnetization proceeds via avalanches triggered by the
increase of the external magnetic field.
Previous studies~\cite{Sethna93,Dhar97} have shown that in the
presence of strong disorder all avalanches are bounded and the
hysteresis loop is macroscopically smooth. In weak disorder,
however, there is a finite coercive field beyond which the
ferromagnetic coupling induces macroscopic avalanches that span
large parts of the sample and lead to a sharp jump in the
hysteresis curve. The two disorder regimes are separated by a
critical point which is believed to control the power law
distribution of Barkhausen
noise~\cite{DahmenSethna96,PerkovicDahmen99Vives03}. Recently,
indications for such a disorder induced transitions have been
reported in disordered magnets~\cite{Berger00,Marcos03} as well as
in the context of capillary condensation in
aerogels~\cite{Tulimieri99,DetcheverryKierlik03}.

The hysteresis in the RFIM is intimately tied to the complexity of
its free energy landscape, possessing an exponential number of
metastable minima. 
However, the precise
connection between the disorder induced energy landscape and the
nature and extent of the avalanches has not been established. So
far, studies of the hysteresis in the RFIM concentrated on lattice
simulations at zero
temperature~\cite{Sethna93,PerkovicDahmen99Vives03}, and most
analytical insight was restricted to the Bethe
lattice~\cite{Dhar97} or the extreme mean field
limit~\cite{Sethna93} where all spins interact with each other. In
this Letter, we make a first step towards an analytical theory for
3D systems. In particular, we analyze the features of the
hysteresis loop as a function of disorder strength and sample size and characterize the typical avalanches occurring along the
hysteresis loop. This provides a new perspective on the
above-mentioned disorder induced transition. Further, by studying
typical configurations that trigger avalanches at zero field we
obtain insight into metastability in the
ferromagnetic phase.

Before giving the details of our analysis, we summarize the
qualitative picture we have obtained, concentrating on the raising
branch of the saturation hysteresis loop (see Fig.1). In a pure
system, hysteresis arises due to the metastability of the
negatively magnetized state in external fields $H$ below
the spinodal field $H_{\rm sp}$.
This simple scenario is strongly modified by disorder: Due to rare
 regions with strong positive fields, bubbles of positive magnetization are created
spontaneously at any small $H$. However, they are prevented from spreading
 by disorder induced pinning. The latter 
leads to a
finite coercive field $H_c$ which grows with disorder strength. For
fields below $H_c$, the magnetization curve is determined by
small avalanches, triggered by the increase of $H$.
At $H_c$ sample spanning avalanches occur, while beyond $H_c$ only small regions with strong negative random
fields may resist the invasion of the positively magnetized phase,
finally collapsing under further increase of the field.
With increasing disorder strength, the local avalanches become so
dense that the emerging bubbles already percolate before the
threshold for unlimited  bubble spread, $H_c$, is reached. In this situation, there is
no room for avalanches to spread at $H_c$, and the
hysteresis loop becomes continuous.

In order to study the nature of avalanches in the 3D RFIM, we have
analyzed its continuum version, described by the standard
Landau-Ginzburg free energy
\begin{equation}
\label{freeenergy} \bar{F}=F_0\int d{\bf
x}\left[\frac{1}{2}(\nabla \varphi)^2 + \bar{V}(\varphi)
-\bar{H}\varphi -\bar{h}({\bf x}) \varphi \right].
\end{equation}
Here $\varphi({\bf x})$ is the coarse-grained magnetization whose
tendency towards ferromagnetism is described by the potential
$\bar{V}(\varphi)=\frac{m^2}{2}(\varphi)^2+\frac{g}{4}(\varphi)^4$
with the reduced temperature $m^2 \propto \tau\equiv -(1-T/T_c)$.
$\bar{H}$ is the uniform external magnetic field, and
$\bar{h}({\bf x})$ is a Gaussian random field with zero mean and
variance $\langle \bar{h}({\bf x}) \bar{h}({\bf x'})\rangle =
\bar{\Delta}\; \delta({\bf x-x'})$. The prefactor $F_0$ depends on
the microscopic details and increases with the range of
ferromagnetic interactions.

{\begin{figure} [t]
\epsfxsize=.45 \textwidth
\centerline{\epsfbox{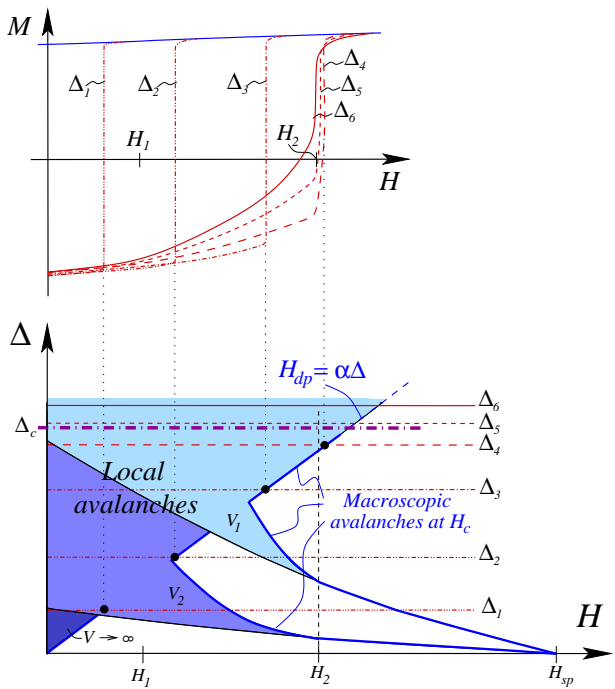}}
\caption{\it Top: \rm Hysteresis curves for increasing disorder.
At the coercive field $H_c$ the magnetization jumps discontinuously. In
strong disorder ($\Delta>\Delta_c\approx \Delta_5$) the jump disappears since
bubbles are already dense at $H_{\rm dp}$. \it Bottom: \rm The shaded
areas indicate where local avalanches are observed in the
$H-\Delta$ plane (in finite volumes $V_1<V_2<\infty$ [lighter
shaded], and for $V=\infty$ [dark area, $H<H_{\rm dp}=\alpha \Delta$]).
Macroscopic avalanches occur at the coercive field $H_c$ (thick blue
line). The horizontal lines correspond to the hysteresis curves in the top
panel.} \label{FigHysteresisloop}
\end{figure}}

We only consider $T<T_c$ ($m^2<0$) which is necessary to ensure
the presence of metastable states. It is
convenient to rescale all quantities according to natural units,
${\bf r}={\bf x}/\ell$, $\Psi({\bf r})=\varphi({\bf
x})/\varphi_0$, $h({\bf r})=(\ell^2 \bar{h}({\bf
x}))/(2\varphi_0)$, and $H=(\ell^2 \bar{H})/(2\varphi_0)$, where
$\varphi_0=(|m^2|/g)^{1/2}$ is the equilibrium magnetization of
the pure system and $\ell=(2/m^2)^{1/2}\sim\tau^{-1/2}$ is the
mean field correlation length. The reduced disorder strength is
$\Delta=(\bar{\Delta}\ell)/(4\varphi_0^2)\sim
(\bar{\Delta})\tau^{-3/2}$. Further, we denote by $\Psi_{\pm}(H)$
the minimum with positive/negative magnetization of
$V(\Psi)-H\Psi$, where $V(\Psi)=-\frac{1}{2} \Psi^2+\frac{1}{4}
\Psi^4$.

In order to study the free energy landscape of this model, we
restrict ourselves to a mean field analysis and concentrate on
local extrema of (\ref{freeenergy}), satisfying
\begin{eqnarray}\label{MFE}
\frac{\delta F}{\delta \Psi({\bf
r})}=-\frac{1}{2}\nabla^2\Psi({\bf r})+V'(\Psi({\bf r}))-H- h({\bf
r})=0.
\end{eqnarray}
In the spirit of the rate independent hysteresis approximation
\cite{Bertotti} we focus on spontaneously occurring events and
neglect thermal activation. This is justified sufficiently far
below  $T_c$ where typical free energy barriers between local
minima are larger than $T$~\cite{noteGinzburg}.

We may restrict the analysis of the hysteresis loop to the raising
branch where the magnetization is everywhere close to $\Psi_-(H)$
at sufficiently negative $H$. As $H$ is increased, the
magnetization either remains in its local free energy minimum,
responding paramagnetically to the field change, or it becomes
locally unstable, and a bubble with higher magnetization is
created spontaneously. This requires a sufficiently strong local
excursion $h({\bf r})$ of the  random field. For weak disorder
such configurations are rare, their density scaling as $\rho
\propto \exp[-S]$, where $S=\int d{\bf r} [h({\bf r})]^2/2\Delta
$. Hence, we may think of them as dilute islands each of which can
be analyzed separately. Further, we may assume that at some
distance from a rare excursion, the random fields are weak such
that the magnetization tends to $\Psi_-$.

The fate of a spontaneously emerging bubble depends on its size.
We will see that {\bf \it typical} bubbles are small and do not spread much
since they are held back by a surface tension barrier. However,
this does not apply to larger (and hence exponentially rarer)
bubbles. Instead, their dynamics is governed by the competition
between the external field exerting a driving force $2H$ on the
surface of the bubble, and the random field disorder, which tends
to pin it. Collective pinning theory predicts the critical field
for domain wall depinning to scale as  $H_{\rm dp}=\alpha\Delta$
\cite{Villain84Bruinsma84}. From a high velocity
expansion~\cite{Leschhorn97} we obtained the estimate
$\alpha\approx 0.055$, which we confirmed by numerical
simulations~\cite{RossoKrauth02}. For  $H<H_{\rm dp}$, all avalanches are
bounded, and the hysteresis loop is dominated by typical, i.e.,
most abundant, small avalanches. Beyond $H_{\rm dp}$, the \it
first \rm unstable bubble which is not constrained by surface
tension induces a sample spanning avalanche and a macroscopic jump
in magnetization, however low the probability per unit volume for
such an event may be. In the thermodynamic limit, the coercive field $H_c$ thus coincides with the depinning field $H_{\rm dp}$. We will discuss mesoscopic fluctuations of $H_c$ in weakly disordered small samples further below. The approach to complete saturation beyond
$H_c$ is governed by small bubbles of negative
magnetization that spontaneously implode upon an increase of
field.

Let us now analyze in more detail the properties of typical avalanches
as a function of disorder and field, both for $H<H_c$ and $H>H_c$.
An avalanche is triggered spontaneously when a local free energy minimum (with
magnetization $\Psi(r)$) becomes unstable upon increase of $H$.
This occurs when  $\Psi(r)$ merges with an adjacent saddle point,
$\Psi_{\rm sp}(r)$, the magnetization difference $\delta \Psi
=\Psi_{\rm sp}-\Psi$ turning into a soft mode of the
Hessian,
\begin{equation}
\label{marginality}
 \left(-1/2\,\nabla^2+V''(\Psi)\right)\delta
\Psi=0.
\end{equation}
Let us denote by ${\cal M}$ the set of all states $\Psi$, such
that there is a soft mode $\delta \Psi$ satisfying
Eq.~(\ref{marginality}). These states are ``marginal'' in the
sense that they are unstable with respect to the smallest
perturbations.

Typical avalanches are triggered by the most probable
configurations (``instantons'') of random fields, $h=h_{\rm
inst}({\bf r}|H')$, such that the current magnetization state,
$\Psi({\bf r})$, becomes marginal ($\Psi \in {\cal M}$) when the external field reaches
$H'$. These instantons thus correspond to non-trivial local minima of
the action
\begin{eqnarray}
\label{functionalPsi}
 S=\frac{1}{2\Delta}\int d{\bf r} h^2 ({\bf
r})= \frac{1}{2\Delta}\int d{\bf r}
\left[\frac{1}{2}\nabla^2\Psi-V'(\Psi)+H\right]^2
\end{eqnarray}
within the set of marginal states $\Psi\in {\cal M}$. Here, we
used Eq.~(\ref{MFE}) to express the random field in terms of
$\Psi$.
Assuming spherically symmetric instantons, the Euler-Lagrange
equation takes the form
\begin{eqnarray}\label{EulerLagrange}
\left[\frac{-\Delta^{(d)}}{2}+V''(\Psi)\right]&\left(\frac{\Delta^{(d)}\Psi}{2}-V'(\Psi)+H\right)&\equiv
\\
\left[\frac{-\Delta^{(d+2)}}{2}+V''(\Psi)\right]&\left(\frac{\Delta^{(d-2)}\Psi}{2}-V'(\Psi)+H\right)
&=Cn_{\Psi}\nonumber
\end{eqnarray}
where $\Delta^{(d)}=\partial_r^2+\frac{d-1}{r}\partial_r$ denotes
the radial part of the Laplacian in $d$ dimensions, $C$ is a
Lagrange multiplier, and $n_\Psi$ is the local normal to the
manifold ${\cal M}$.

Performing an extensive search for local minima of $S$
 (analytically by solving (\ref{EulerLagrange}) and numerically by
variational minimization of (\ref{functionalPsi})) we found two types of optimal random
field configurations. For $H< H_1=0.0763$, the optimal random
field is such that the state $\Psi$ corresponds to a threefold
degenerate free energy minimum. However, slight deviations $\delta
h(r)$ from the instanton configuration $h_{\rm inst}$ lead to a
local free energy landscape where the state with lowest
magnetization is marginally unstable with respect to a nearby
minimum with locally higher magnetization ($\delta M \propto (\int
\delta h^2)^{1/4}$). An essentially identical analysis for $H_c<H<
+\infty$, but with inverse boundary conditions [$\Psi_-
\rightarrow \Psi_+$], yields similar instantons corresponding to
the dominant bubbles which implode upon increase of the external
field beyond $H_c$~\cite{MullerSilva05}.

A second type of instantons describes the \it typical \rm
avalanches in the interval $H_1<H<H_c$. These instantons can be found
analytically by solving the ``dimensionally reduced'' equations of
motion, $\label{dimred} -1/2\,\Delta^{(d-2)}\Psi+V'(\Psi)-H=0$,
which indeed yield a solutions of Eq.~(\ref{EulerLagrange}) with
$C=0$. The corresponding random field is $h_{\rm inst}(r|
H)=-(\partial_r\Psi)/r$. In $d=3$ dimensions, this equation is
readily integrated with boundary conditions $\Psi'(r=0)=0$ and
$\Psi(r \rightarrow\infty) \rightarrow \Psi_{-}(H)$. The action
Eq. (\ref{functionalPsi}) associated to these field configurations
decreases monotonically from $S(H_1)=5.7 /\Delta$ to $S(H_{\rm
sp})=0$.

{\begin{figure}
\epsfxsize=6.5cm
\centerline{\epsfbox{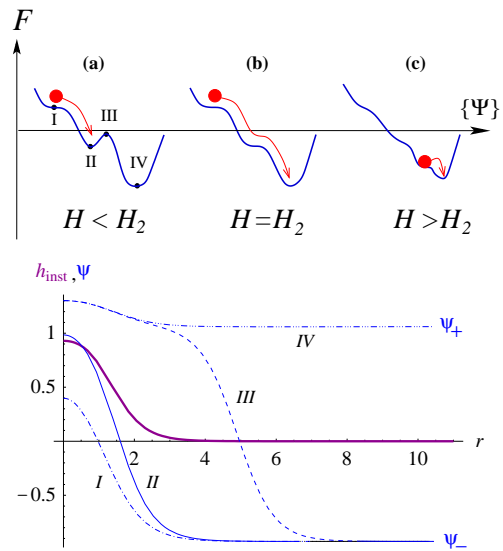}} \vspace{0.5cm} \epsfxsize=7cm
\caption{\textit{Top:} Schematic sketch of the free energy
landscapes associated with typical avalanches for various regimes
of the external field $H$. (a) $H<H_2$ (cf. bottom panel): The
state (I) becomes unstable and runs to the metastable (II) with a
local bubble of higher magnetization. It is separated from the
global minimum (IV) by a surface tension barrier (III); (b)
$H=H_2$: state (II) and (III) merge and the surface barrier
vanishes; (c) $H>H_2$: Rare bubbles in negative random fields
collapse under further increase of $H$. \textit{Bottom:} Solutions
(I-IV) of Eq.~(\ref{MFE}), obtained numerically for $H=0.13<H_2$
and $h=h_{\rm inst}(r|H)$ [thick line].} \label{Figlandscape}
\end{figure}}

An analysis of the local free energy landscape associated to this
type of instantons shows that for
$H_1<H<H_2=0.173$ the typical bubbles are constrained by a surface
tension barrier, the change of magnetization remaining local (see Fig.~\ref{Figlandscape}(a)). This
is illustrated in the lower panel of Fig.~\ref{Figlandscape} where we plot
the instanton $h_{\rm inst}(r|H)$ together with all solutions of
Eq.~(\ref{MFE}) for $H=0.13$. Initially, the system is in the
marginal state (I) with the lowest magnetization. Upon slight
increase of $H$, it becomes unstable and the system runs to the
nearby lower lying minimum (II) which represents a localized
bubble of positive magnetization. Further spread of this bubble is
prevented by the surface tension of the domain wall that separates
the positively and negatively magnetized regions. Indeed, a
further saddle point (III) separates the local minimum from the
global minimum (IV). Note that as long as $H<H_{\rm dp}$,  domain wall pinning by \it typical
\rm random fields will split the state (IV) into a set of
metastable states which differ on larger scales which are not
shown in Fig.~\ref{Figlandscape}.

Beyond $H_2$, the confining surface tension barrier
vanishes, and the avalanches spread further out (cf. Fig.~\ref{Figlandscape}(b,c)), leading to a
marked jump in the slope of the hysteresis curve. However, this
effect is only observable in a small window of disorder (see
Fig.~\ref{FigHysteresisloop}). Indeed, for $\Delta<H_2/\alpha$, macroscopic
avalanches set in before $H_2$, while for
$\Delta\gtrsim H_2/\alpha$ the density of bubbles becomes rapidly
high, preventing large avalanches as discussed below.

At a critical disorder $\Delta_c$ the typical bubbles start to
percolate at the depinning field $H_{\rm dp}$ (see
Fig.~\ref{FigHysteresisloop}). This marks the crossover towards a
regime with a continuous hysteresis loop. Indeed, for
$\Delta>\Delta_c$ bubbles are dense already below the depinning
threshold  $H_{\rm dp}$, such that avalanches at  $H_{\rm dp}$ run into already
positively magnetized regions and thus die out. We can estimate
$\Delta_c$ from the requirement that the density of typical
bubbles is  $\rho(H_{\rm dp}[\Delta_c])={\cal O}(1)$, or $S(H_{\rm dp}[\Delta_c])
\simeq 1$. A more careful calculation including Gaussian
prefactors yields $\Delta_c \simeq 3.2$, which translates into
$\bar{\Delta}_c \simeq 9 \;(|m^2|)^{3/2}/g \simeq const\times
\tau^{3/2}/g$  for the transition line in the $T-\Delta$ plane. It
is interesting to note that for $\Delta\approx\Delta_c$ the depinning field 
$H_{\rm dp}=\alpha \Delta$ is very close to the threshold field $H_2$ for
typical avalanches to overcome surface tension barriers.

Above we used the identity of the coercive field
and the depinning threshold. This applies to the infinite volume
limit, as a result of which one finds $H_c[\Delta\rightarrow 0]
\rightarrow 0$ (see, e.g., \cite{Shukla}), instead of
$H_c[\Delta\rightarrow 0] \rightarrow H_{\rm sp}$, which holds for
finite size systems (see bottom of Fig.~\ref{FigHysteresisloop}),
$H_{\rm sp}=2/3\sqrt{3}\approx 0.385$ being the spinodal field of
the pure system. For a finite system of volume $V$, the only type of
bubbles occurring in the hysteresis loop are those with
sufficiently large density, $V\rho(H,\Delta) \gtrsim 1$. This
relation allows us to estimate the field $H^*$ at which the first
(typical) bubbles occur. If the disorder is so weak that
$H^*>H_2$, there will be a single sample spanning avalanche around
the coercive field  $H^* \simeq H_{\rm sp}-0.5 \Delta
\log[V/\xi^3]$. For stronger disorder ($H^*<H_2$) local bubbles
occur at $H^*$ while only at higher field a macroscopic avalanche
is triggered by the first bubble that is not constrained by surface tension.

 {\begin{figure} \epsfxsize=4cm
\centerline{\epsfbox{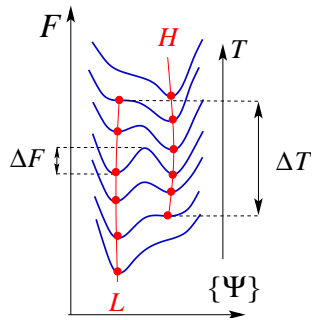}} \vspace*{0cm}
\caption{Schematic view of the temperature evolution of the local
free energy landscape in a random field $h_{\rm
inst}(r|H=0)+\delta h$. Upon cooling, the high temperature state
($H$) becomes unfavorable with respect to an emerging low energy
state with more uniform magnetization ($L$), but remains
metastable in a temperature range $\Delta T$ due to a free energy
barrier $\Delta F$.} \label{Fig3}
\end{figure}}

The analysis of instantons also provides information on the
emergence and bifurcation of metastable states upon temperature
variations. We have studied the temperature evolution of the free
energy landscape in the presence of local random field
configurations close to the instantons corresponding to $H=0$ and
$T=T_0$, $h=h_{\rm inst}(H=0;T_0)+\delta h$. These describe the
rare regions which locally admit two metastable configurations in
the ferromagnetic phase, cf.~Fig.~\ref{Fig3}. At high temperatures
there is only one state ($H$) in which $\Psi(r)$ adjusts to the
random field. At lower temperature, a second state ($L$) with more
uniform magnetization emerges and eventually dominates. This
confirms a scenario for the emergence of metastability conjectured
by Villain in the context of pinned domain
walls~\cite{Villain84Bruinsma84}. Over a range of order $\Delta
T\sim \delta^{3/2}$ (where $\delta\equiv(\int \delta h^2)^{1/2}$)
the two states coexist, being separated by a free energy barrier
of order $\Delta F\sim \delta^2$. The spatial density of such
two-level systems with a given free energy barrier scales as
$\rho(\Delta F) \propto  \exp\left[-S_0-O(\Delta F^{1/2})\right]$
with $S_0=S[h_{\rm inst}(r|H=0)]\simeq 8.2 /\Delta$. These two-level
systems are very similar to the metastable bubbles observed
numerically {\it above} the ferromagnetic
transition~\cite{Lancaster95}. We expect a deeper analysis of the
bifurcation of states along the lines of this Letter to yield
new insight into the nature of the phase transition in the RFIM.

We thank J. Cardy, T. Nattermann,
M. Rosinberg, G. Tarjus, and in particular, L. B. Ioffe for useful
discussions. We are grateful to  A. Rosso for providing us the code of
\cite{RossoKrauth02}. This work was supported by NSF grant
DMR-0210575.

\end{document}